\documentclass[twocolumn,showpacs,preprintnumbers,amsmath,amssymb,prl,superscriptaddress]{revtex4}
\usepackage{graphicx}% Include figure files
\usepackage{dcolumn}% Align table columns on decimal point
\usepackage{bm}% bold math

\begin{document}

\title{Precision study of the \boldmath{$dp{\to}\,^{3}$He$\,\eta$} reaction for excess energies between 20 and 60~MeV}

\author{ T.\,Rausmann}\email[E-mail: ]{trausmann@uni-muenster.de}
\affiliation{Institut f\"ur Kernphysik, Universit\"at
M\"unster, D-48149 M\"unster, Germany}%
\author{A.\,Khoukaz}%\email[E-mail: ]{khoukaz@uni-muenster.de}
\affiliation{Institut f\"ur Kernphysik, Universit\"at
M\"unster, D-48149 M\"unster, Germany}%
\author{M.\,B\"{u}scher}
\affiliation{Institut f\"ur Kernphysik and J\"ulich Centre for Hadron Physics, Forschungszentrum
J\"ulich, D-52425 J\"ulich, Germany}
\author{D.\,Chiladze}
\affiliation{Institut f\"ur Kernphysik and J\"ulich Centre for Hadron Physics, Forschungszentrum
J\"ulich, D-52425 J\"ulich, Germany}
\affiliation{High Energy Physics Institute, Tbilisi State
University, 0186 Tbilisi, Georgia}
\author{S.\,Dymov}
\affiliation{Physikalisches Institut II, Universit{\"a}t
Erlangen-N{\"u}rnberg, D-91058 Erlangen, Germany }
\affiliation{Laboratory of Nuclear Problems, JINR, RU-141980 Dubna,
Russia}
\author{M.\,Hartmann}
\affiliation{Institut f\"ur Kernphysik and J\"ulich Centre for Hadron Physics, Forschungszentrum
J\"ulich, D-52425 J\"ulich, Germany}
\author{A.\,Kacharava}
\affiliation{Institut f\"ur Kernphysik and J\"ulich Centre for Hadron
Physics, Forschungszentrum J\"ulich, D-52425 J\"ulich, Germany}
\author{I.\,Keshelashvili}
\affiliation{Physics Dept., University of Basel, Klingelbergstrasse 82, 4056 Basel, Switzerland}
\author{P.\,Kulessa}
\affiliation{H.\,Niewodniczanski Institute of Nuclear Physics PAN,
PL-31342 Cracow, Poland}
\author{Y.\,Maeda}
\affiliation{Research Center for Nuclear Physics, Osaka University, Ibaraki, Osaka 567-0047, Japan}
\author{T.\,Mersmann}
\affiliation{Institut f\"ur Kernphysik, Universit\"at
M\"unster, D-48149 M\"unster, Germany}%
\author{M.\,Mielke}%
\affiliation{Institut f\"ur Kernphysik, Universit\"at
M\"unster, D-48149 M\"unster, Germany}%
\author{S.\,Mikirtychiants}
\affiliation{High Energy Physics Department, Petersburg Nuclear Physics Institute, RU-188350 Gatchina, Russia}
\author{M.\,Nekipelov}
\affiliation{Institut f\"ur Kernphysik and J\"ulich Centre for Hadron
Physics, Forschungszentrum J\"ulich, D-52425 J\"ulich, Germany}
\author{H.\,Ohm}
\affiliation{Institut f\"ur Kernphysik and J\"ulich Centre for Hadron Physics, Forschungszentrum
J\"ulich, D-52425 J\"ulich, Germany}
\author{M.\,Papenbrock}%
\affiliation{Institut f\"ur Kernphysik, Universit\"at
M\"unster, D-48149 M\"unster, Germany}%
\author{F.\,Rathmann}
\affiliation{Institut f\"ur Kernphysik and J\"ulich Centre for Hadron Physics, Forschungszentrum
J\"ulich, D-52425 J\"ulich, Germany}
\author{V.\,Serdyuk}
\affiliation{Laboratory of Nuclear Problems, JINR, RU-141980 Dubna,
Russia}
\author{H.\,Str\"oher}
\affiliation{Institut f\"ur Kernphysik and J\"ulich Centre for Hadron
Physics, Forschungszentrum J\"ulich, D-52425 J\"ulich, Germany}
\author{A.\,T\"{a}schner}
\affiliation{Institut f\"ur Kernphysik, Universit\"at
M\"unster, D-48149 M\"unster, Germany}%
\author{Yu.\,Valdau}
\affiliation{High Energy Physics Department, Petersburg Nuclear Physics Institute, RU-188350 Gatchina, Russia}
\author{C.\,Wilkin}
\affiliation{Physics and Astronomy Department, UCL, Gower Street,
London WC1E 6BT, UK}%
\date{\today}

\begin{abstract}
The differential and total cross sections for the
$dp\to\,^{3}\textrm{He}\,\eta$ reaction have been measured at
COSY--ANKE at excess energies of 19.5, 39.4, and 59.4~MeV over the
full angular range. The results are in line with trends apparent from
the detailed near-threshold studies and also agree with those from
CELSIUS, though the present data have higher precision. While at
19.5~MeV the results can be described in terms of $s$- and $p$-wave
production, by 59.4~MeV higher partial waves are required. Including
the 19.5~MeV point together with the near-threshold data in a global
$s$- and $p$-wave fit gives a poorer overall description of the data
though the position of the pole in the $\eta^3$He scattering
amplitude, corresponding to the quasi-bound or virtual state, is
hardly changed.
\end{abstract}

 \maketitle
%%%%%%%%%%%%%%%%%%%%%%%%%%%%%%%%%%%%%%%%%%%%%%%%%%%%%%%%%%%%%%%%%

We have recently provided results on the
$dp\to\,^{3}\textrm{He}\,\eta$ reaction  very near threshold in fine
energy steps. These show that the total cross section reaches a
plateau for an excess energy $Q$ that is less than
1~MeV~\cite{Timo07} and this behavior was confirmed by an independent
measurement~\cite{Smy07}. The abrupt variation of the cross section
can be understood~\cite{Wil93} as being due to a final state
interaction (FSI) that is enhanced through the presence of a
quasi-bound or virtual state pole in the $\eta^3$He elastic
scattering amplitude for $|Q|<1$~MeV, though its location in the
complex planes is ambiguous.

Further evidence for the pole hypothesis is to be found from the
study of the angular distribution~\cite{Wil07}. For $Q<11$~MeV a
linear dependence on the cosine of the $\eta$ production angle is
seen, which suggests that only the $\eta$ $s$-wave and $s$-$p$
interference are important in this region. However, the energy
dependence of the angular slope shows that the phase of the
interference changes rapidly with $Q$, presumably due to the pole in
the $s$-wave production amplitude. It is therefore of interest to
investigate how this behavior extends to higher energies and this we
have done through measurements at $Q=19.5$, 39.4, and 59.4~MeV. The
first two energies overlap with results obtained by a CELSIUS
collaboration~\cite{Bil02} and, while the data sets are largely
consistent, the present ones allow firmer conclusions to be drawn.
The energy range 10 -- 60~MeV seems to be a transition region going
from one where there is $s+p$ dominance to a regime where higher
partial waves become important.

\begin{figure}[hbt]
\includegraphics[width=8cm]{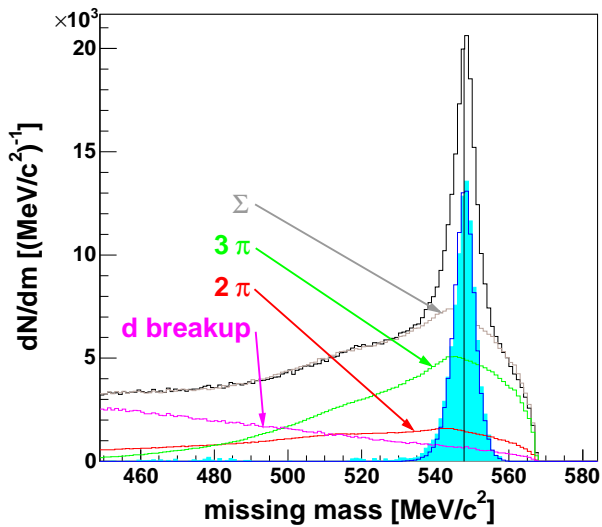}
\caption{(Color online) Missing-mass distribution for the
$dp\to\,^{3}\textrm{He}\,X$ reaction at an excess energy of 19.5~MeV
with respect to the $\eta$ threshold. Contributions of simulated
reactions to the background fit and their sum ($\Sigma$) are shown.
After subtracting these from the data, the difference histogram shows
a clean $\eta$ peak (shaded) that agrees very well with the
simulation (solid histogram) of the $dp\to\,^{3}\textrm{He}\,\eta$
reaction.\label{MM}}
\end{figure}

The experiment was carried out using the ANKE
spectrometer~\cite{ANKE} in combination with a hydrogen cluster
target~\cite{target} at an internal station of the Cooler Synchrotron
COSY--J{\"u}lich. The conditions were identical to those of our
near-threshold work~\cite{Timo07} with the exception that, instead of
using a beam that was continuously ramped in energy, three fixed
values of the beam momentum were requested, \emph{viz} 3.223, 3.306,
and 3.389~GeV/$c$. The produced $^{3}\textrm{He}$ were detected in
the ANKE forward detection system, which consists of one drift
chamber, two multi-wire proportional chambers, and three layers of
scintillation hodoscopes. The trigger used demanded hits in the first
two layers. The tracks of charged particles could be traced back
through the precisely known magnetic field to the interaction point,
leading to a momentum reconstruction for registered particles. The
$^3$He were then identified by making a cut in the energy loss
\emph{versus} momentum plot using the information from all the
hodoscope layers.

The missing-mass distribution for all $dp\to\,^{3}\textrm{He}\,X$
events measured at $Q=19.5$~MeV is presented in Fig.~\ref{MM}. This
shows a prominent $\eta$ peak sitting on a background that is rather
similar to that observed in the below-threshold data taken under
identical conditions to the ones used here~\cite{Timo07}. The
spectrum has been modeled in a phase-space Monte Carlo simulation of
\mbox{two-,} three-, and four-pion production plus a small component
arising from misidentified protons from the intense deuteron breakup
reaction. These simulations were fitted together with that for the
$dp\to\,^{3}\textrm{He}\,\eta$ reaction to the data. After
subtracting the sum of the background reactions from the measured
points, a clean $\eta$ peak was found.

In order to determine the differential cross section for each excess
energy, the whole range of the $^{3}\textrm{He}$ c.m. production
angles was divided into 20 bins and a missing-mass distribution
constructed for each of them. The $\eta$ content was determined in a
similar manner to that shown in Fig.~\ref{MM}.

In the near-threshold experiment the value of $Q$ could be determined
directly from the data by studying the size of the momentum
ellipse~\cite{Timo07}. This leads to significant errors as $Q$ gets
larger and the ellipse expands. Therefore, the excess energy was
calculated from the preset COSY beam momentum. This leads to an
uncertainty of 0.8~MeV, which is consistent with the precision of
$0.1\%$ in the COSY beam momentum. For excess energies in the 20 --
60~MeV range this uncertainty is of minor importance.

Just as in the earlier work~\cite{Timo07}, the luminosity
$\mathcal{L}$ needed to convert counts to cross sections was found
through the simultaneous measurement in the forward detector of the
deuteron from $dp$ elastic scattering. The cross section varies very
fast with the deuteron angle in this region~\cite{dpelastic} and it
is the systematic uncertainty in the determination of this angle that
dominates the error in $\mathcal{L}$ of about $\pm15$\%. However, it
is important to stress that this generally affects all three energies
in the same way, as it does also the near-threshold
data~\cite{Timo07}.

\begin{figure}[hbt]
\includegraphics[scale=0.4]{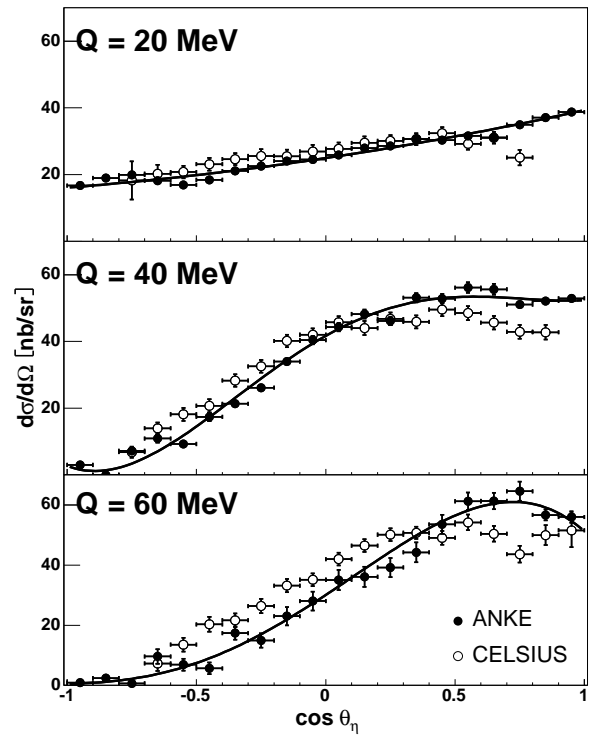}
\caption{Differential cross sections for the three excess energies
studied at ANKE (filled circles). The CELSIUS data (open circles)
shown in the 60~MeV plot were measured at 80~MeV~\cite{Bil02}. The
solid lines represent fits of Eq.~(\ref{polynomial}) to the ANKE data,
with the parameters being given in
Table~\ref{tab:table1}.\label{dcs}}
\end{figure}

\begin{table}[hbt]
\caption{\label{tab:table1} Results for the
$dp\to\,^{3}\textrm{He}\,\eta$ differential and total cross sections.
In addition to the statistical errors, there is a common systematic
uncertainty of 15\% in the total cross sections, due mainly to the
luminosity determination. The fit parameters $a_n$ of
Eq.~(\ref{polynomial}) are similarly affected by this scale
uncertainty.}
\begin{ruledtabular}
\begin{tabular}{cccc}
$Q$ [MeV]& $19.5\pm 0.8$ & $39.4\pm 0.8$ & $59.4\pm 0.8$\\
$p_{\eta}$ [MeV/$c$]& $134.5\pm2.8\phantom{1}$ & $192.4\pm2.0\phantom{1}$ & $237.7\pm1.6\phantom{1}$\\
\hline
$a_0$ [nb/sr] & $24.9\pm0.3$ & \phantom{-}$41.7\pm0.9$ & \phantom{-}$30.1\pm1.5$\\
$a_1$ [nb/sr]&  $11.6\pm0.4$ & \phantom{-}$43.4\pm1.8$ & \phantom{-}$57.0\pm3.2$\\
$a_2$ [nb/sr]&  $\phantom{1}2.8\pm0.8$ & -$36.7\pm4.6$ & \phantom{-}$11.5\pm7.4$ \\
$a_3$ [nb/sr]& \phantom{--}-- & -$18.6\pm2.2$ & -$31.7\pm4.5$\\
$a_4$ [nb/sr]& \phantom{--}-- & $\phantom{0}22.6\pm4.5$  & -$15.7\pm8.0$ \\
\hline $\sigma_{\text{tot}}$ [nb] & $326.7\pm2.0$ & $428.8\pm3.4$ &
$388.1\pm7.2$
\end{tabular}
\end{ruledtabular}
\end{table}

\begin{figure*}[hbt]
\includegraphics[width=17cm]{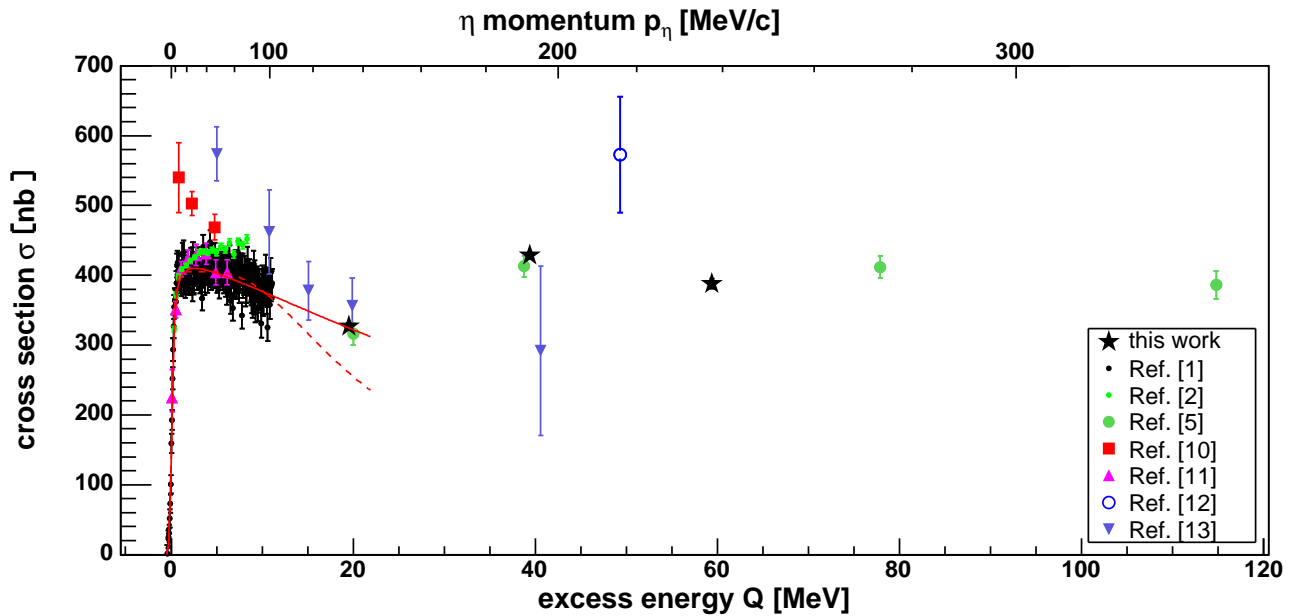}
\caption{(Color online) Comparison of the total cross sections
obtained in this experiment (black stars) with previous data:
Ref.~\cite{Timo07} (small black circles), Ref.~\cite{Smy07} (small
green circles), Ref.~\cite{Bil02} (large green circles),
Ref.~\cite{Ber88} (large red squares), Ref.~\cite{May96} (magenta
triangles), Ref.~\cite{Bet00} (open blue circle), Ref.~\cite{Adam07}
(inverted blue triangles). The solid and dashed lines show the result
of the recursive fit to the data with and without considering the 19.5~MeV
point.
\label{sigt}}
\end{figure*}

Figure~\ref{dcs} shows the angular distributions obtained at the
three different energies. Also presented are the points measured by a
CELSIUS collaboration in the vicinity of $Q=20$~MeV and 40~MeV, as
well as those at 80~MeV~\cite{Bil02}. These are generally in
agreement with the present results, though our data have smaller
statistical error bars and cover the complete $\cos\theta_{\eta}$
range. It is important to note that there is no sign of a forward dip
at 19.5~MeV and that any at 39.4~MeV is much weaker than the one
found at CELSIUS~\cite{Bil02}. The curves show polynomial fits to the
ANKE points
\begin{equation}
\label{polynomial}
\frac{d\sigma}{d\Omega}=\sum_{n=0}^{4}a_n\left(\cos\theta_{\eta}\right)^n\,,
\end{equation}
with the values of the parameters $a_n$ being given in
Table~\ref{tab:table1}. These prove that, although it might just be
sufficient to retain only $s$ and $p$ waves for the 19.5~MeV data,
$d$ and higher waves are required to describe the 39.4 and 59.4~MeV
data. The negative value of $a_4$ at 59.4~MeV indicates that at least
$f$ waves are needed here.

The values obtained for the total $dp\to\,^{3}\textrm{He}\,\eta$
cross sections are also reported in Table~\ref{tab:table1} and
compared with other published results in Fig.~\ref{sigt}. Not shown
there is the 15\% uncertainty in the ANKE data that arises mainly
from the luminosity determination. As discussed earlier, this is
largely a common factor that does not affect the discussion of the
energy dependence. The rise apparent between 20 and 40~MeV in both
our data and those of CELSIUS~\cite{Bil02} possibly reflects the
increased influence of the higher partial waves that are needed to
describe the angular dependence of Fig.~\ref{dcs}.

\begin{figure}
\includegraphics[scale=0.4]{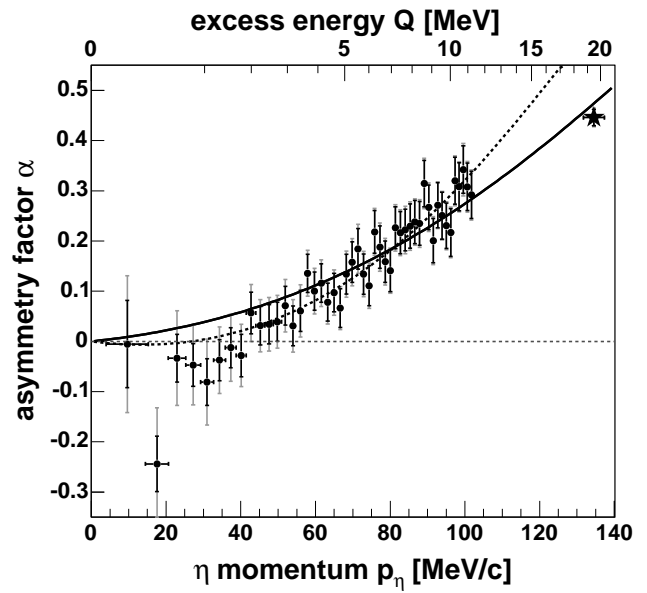}
\caption{Slope parameter $\alpha$ of Eq.~(\ref{alpha}) as a function
of the $\eta$ c.m.\ momentum. Bin widths and statistical errors are
shown bold; systematic uncertainties are shown with feint lines. The
solid and dashed lines show the result of the recursive fit to the
data with and without considering the new data point at
$Q=19.5$~MeV.\label{slope}}
\end{figure}

The strong forward/backward asymmetry shown by the data of
Fig.~\ref{dcs} arises from the interference between odd and even
partial waves and at low energies this can be summarized by a slope
parameter, defined by
\begin{equation}
\label{alpha} \alpha =\left. \frac{d}{d(\cos\theta_\eta)}\ln\left(
\frac{d\sigma}{d\Omega}\right)\right|_{\cos\theta_\eta=0}\,\cdot
\end{equation}
The values of $\alpha$ deduced at 19.5~MeV and from our previous
measurements are shown in Fig.~\ref{slope}. The results presented in
Ref.~\cite{Smy07} show a very similar behavior, as do those of
Ref.~\cite{May96}, though with much lower precision.

Since the 19.5~MeV differential cross section can be fit by a
quadratic in $\cos\theta_{\eta}$, these data might be described in
terms of $s$- and $p$-wave production amplitudes. It is therefore of
interest to try to include these values together with our
near-threshold measurements of the reaction~\cite{Timo07} in a global
energy-dependent fit to see if the $s$+$p$ hypothesis holds up to this
energy.

As pointed out in Ref.~\cite{Wil07}, due to the spin complexity,
there are two independent $s$-wave production amplitudes and five
$p$-wave terms. The FSI should affect the two $s$-wave terms in
similar ways and some support for this is to be found in the weak
energy dependence of the deuteron tensor analyzing powers of the
reaction~\cite{Ber88}. The ansatz of Ref.~\cite{Wil07} therefore
assumes average $s$- and $p$-wave amplitudes, $f_s$ and
$p_{\eta}f_p$, in terms of which the total cross section and slope
parameter are expressed as
\begin{eqnarray}
\label{stot_amps} \sigma_{\text{tot}}&=&4\pi\,\frac{p_{\eta}}{p_p}
\left[|f_{s}|^2+p_{\eta}^2|f_{p}|^2\right]\,\\
\label{alpha_amps} \alpha&=&2p_{\eta}\frac{\textit{Re}(f_{s}^*f_{p})}
{|f_{s}|^2+p_{\eta}^2|f_p^2|}\,,
\end{eqnarray}
where $p_p$ is the initial proton c.m\ momentum.

The rapid rise in the total cross section from threshold is due to
the pole in the $s$-wave amplitude $f_s$ and this is also responsible
for the non-linear behavior of the slope parameter $\alpha$ with the
$\eta$ c.m.\ momentum shown in Fig.~\ref{slope}. This comes about
because of the variation of the phase of the interference
$\textit{Re}(f_{s}^*f_{p})$ with $p_{\eta}$ in
Eq.~(\ref{alpha_amps}).

The $s$-wave amplitude was taken as the product of a near and distant
pole,
\begin{equation}
f_{s}=\frac{f_B}{(1-p_{\eta}/p_1)(1-p_{\eta}/p_2)}\,,
\end{equation}
where the distant one is an effective pole that absorbs any
residual momentum dependence so that $f_B$ is taken as
constant~\cite{Timo07}, as is the reduced $p$-wave amplitude $f_p$.

Before the ansatz can be compared to the experimental data, it has to
be smeared over the spread in the initial deuteron beam momentum
inside COSY~\cite{Timo07}. Fitting the resulting formulae to the
total cross section and asymmetry data leads to the dashed curves in
Figs.~\ref{sigt} and \ref{slope} if the 19.5~MeV point is not
included in the fit, as it was not in our near-threshold
work~\cite{Timo07}. Including this point leads to the solid curves,
which gives a much poorer description of the asymmetry at low
$p_{\eta}$, though the differences are much harder to see for the
total cross section in the near-threshold region.

Despite the poorer overall description of the low energy data, the
position of the nearby pole in the complex energy plane is only
changed from $|Q|\approx 0.4$~MeV to $\approx 0.9$~MeV through the
inclusion of the 19.5~MeV point. This shows that the position of the
${\eta}^{\,3}$He pole is robustly fixed in magnitude, though the data
cannot determine whether it is a quasi-bound or virtual state. The
data also indicate that the $s$-wave amplitude carries on decreasing
with energy until at least 19.5~MeV with the $p$-waves becoming
steadily more important. However, the higher $\chi^2$/ndf obtained
when including the 19.5~MeV point (1.61 \emph{versus} 1.42) and the
systematically poorer description of the asymmetry data in
Fig.~\ref{slope} suggest that the modelling of in terms of only
spin-averaged $s$- and $p$-wave amplitudes is insufficient at this
energy.

In summary, we have extended the measurements of the
$dp\to\,^{3}\textrm{He}\,\eta$ differential and total cross sections
to higher excess energies. Although the angular distribution at
19.5~MeV might be described in terms of just $s$ and $p$ waves, a
global fit of the data using the methodology of Ref.~\cite{Wil07}
gives unsatisfactory results and this suggests that higher partial
waves are probably already significant at this energy. The negative
value of the $a_4$ parameter in Table~\ref{tab:table1} is a sign that
at least $f$ waves are important at 59.4~MeV. On the other hand, even
if the fit is forced to describe a data set that includes the
19.5~MeV point, the position of the ${\eta}^{\,3}$He pole hardly
moves.

The analysis of Ref.~\cite{Wil07} assumed that the energy dependence
of the two $s$ waves was identical. This can only be tested through
precise measurements of the deuteron tensor analyzing powers of the
reaction in finer energy steps than are currently
available~\cite{Ber88}. Data on this observable will be provided by
the COSY-ANKE collaboration~\cite{Tob06}.

\begin{acknowledgments}
The authors wish to record their thanks to the COSY machine crew for
producing such good experimental conditions and also to other members
of the ANKE collaboration for diverse help in the experiment. This
work was supported in part by the HGF--VIQCD, and JCHP FFE.
\end{acknowledgments}
%%%%%%%%%%%%%%%%%%%%%%%%%%%%%%%%%%%%%%%%%%%%%%%%%%%%%%%%%%%%%%%%%%%%%%%%%%%%

\end{document}